 \documentclass{emulateapj}

\newcommand{\msun}{$M_{\odot}$}

\newcommand {\Lya}    {Ly$\alpha$}   

\newcommand {\HI}        {\ion{H}{1}}   

\newcommand {\HeII}     {\ion{He}{2}}   

\newcommand {\OVI}    {\ion{O}{6}}   

\newcommand {\CIII}    {\ion{C}{3}}   
\newcommand {\CIV}    {\ion{C}{4}}

\newcommand {\SiIV}   {\ion{Si}{4}}
\newcommand {\SiIII}  {\ion{Si}{3}}

\newcommand {\kms}    {km~s$^{-1}$}

\newcommand {\HST}    {{\it HST}}

\newcommand {\etal}   {et~al.} 

\newcommand{\uvunits} {erg~cm$^{-2}$~s$^{-1}$~Hz$^{-1}$~sr$^{-1}$} 

\begin{document}

\title{The Metagalactic Ionizing Background:  A Crisis in UV Photon Production 
or Incorrect Galaxy Escape Fractions?}  

\author{J. Michael Shull, Joshua Moloney, Charles W. Danforth, \& Evan M. Tilton}
\affil{CASA, Department of Astrophysical \& Planetary Sciences, \\
University of Colorado, Boulder, CO 80309}

\email{michael.shull@colorado.edu,  joshua.moloney@colorado.edu, 
danforth@colorado.edu, evan.tilton@colorado.edu  \\ }  


\begin{abstract} 

Recent suggestions of a ``photon underproduction crisis" (Kollmeier \etal\ 2014) have generated 
concern over the intensity and spectrum of ionizing photons in the metagalactic ultraviolet background 
(UVB).   The balance of hydrogen photoionization and recombination determines the opacity of 
the low-redshift intergalactic medium (IGM).   We calibrate the hydrogen photoionization rate 
($\Gamma_{\rm H}$) by comparing  {\it Hubble Space Telescope} spectroscopic surveys of the low-redshift 
column density distribution of \HI\ absorbers and the observed ($z < 0.4$) mean \Lya\ flux decrement, 
$D_A = (0.014)(1+z)^{2.2}$, to new cosmological simulations.  The distribution,
$f(N_{\rm HI}, z) \equiv d^2 {\cal N} / d(\log N_{\rm HI}) dz$, is consistent with an increased UVB 
that includes contributions from both quasars and galaxies.  Our recommended fit, 
$\Gamma_{\rm H}(z) = (4.6 \times 10^{-14}$ s$^{-1})(1+z)^{4.4}$ for $0 < z < 0.47$, corresponds to
unidirectional  LyC photon flux $\Phi_0 \approx 5700$~cm$^{-2}$~s$^{-1}$ at $z = 0$.  This flux 
agrees with observed IGM metal ionization ratios (\CIII/\CIV\ and \SiIII/\SiIV) and suggests a 25-30\% 
contribution of \Lya\ absorbers to the cosmic baryon inventory.  The primary uncertainties in the low-redshift 
UVB are the contribution from massive stars in galaxies and the LyC escape fraction ($f_{\rm esc}$), a 
highly directional quantity that is difficult to constrain statistically.   
We suggest that both quasars and low-mass starburst galaxies are important contributors to the ionizing 
UVB at $z < 2$.  Their additional ionizing flux would resolve any crisis in photon underproduction. 

\end{abstract} 


\keywords{cosmological parameters --- ultraviolet: galaxies --- observations --- intergalactic medium 
--- quasars:  absorption lines  }

\section{INTRODUCTION}

One the most important but poorest known parameters in studies of the intergalactic medium (IGM) 
and circumgalactic medium (CGM) is the intensity of the metagalactic ionizing ultraviolet background 
(UVB).  This uncertainty is not surprising, considering the strong absorption of extreme ultraviolet (EUV) 
photons at wavelengths $\lambda \leq 911.753$~\AA\  by the interstellar medium (ISM).   Lacking direct 
measurements of the metagalactic radiation field in the Lyman continuum (LyC), astronomers rely on 
indirect probes of photoionizing radiation and theoretical estimates\footnote{UVB models include 
calculations by Haardt \& Madau (1996, 2001, 2012), denoted HM96, HM01 and HM12, Shull \etal\ (1999),
and Faucher-Gigu\`ere \etal\ (2009).  HM05 refers to the higher-flux spectrum in a 2005 August update to 
the Haardt \& Madau (2001)  ``Quasars+Galaxies"  spectra, provided for inclusion in the photoionization 
code \textsc{Cloudy}.  The hydrogen photoionization rates, $\Gamma_{\rm H}$, in these models vary by
up to a factor of six (see Table 1).}  
based on the presumed ionizing sources and cosmological radiative transfer.  An accurate characterization 
of this UVB is necessary for modeling the thermal and ionization conditions (Bolton \etal\ 2014; Boera 
\etal\ 2014) of the IGM and CGM, for a census of IGM baryon content (Shull \etal\ 2012b), and for the 
ionization corrections needed to derive metallicities (Shull \etal\ 2014; Werk \etal\ 2014).  For 
photoionization of hydrogen, helium, and many spectroscopically accessible ions of heavy elements 
(C, N, O, Ne, Mg, Si, S, Fe) the most important spectral range extends from the hydrogen Lyman edge 
(energies $E \geq 13.6$ eV) through the \HeII\ continuum ($E \geq 54.4$~ eV)  and into the extreme 
ultraviolet (EUV) and soft X-ray (100-1000 eV).  
  
The claim of a ``photon underproduction crisis" (Kollmeier \etal\ 2014, hereafter denoted K14) generated 
considerable anxiety in the extragalactic astronomy community and prompted a re-examination of the 
assumptions in modeling the EUV background at low redshift.  These authors compared predictions from 
smoothed-particle hydrodynamic (SPH) simulations to observed properties of low-redshift  \HI\ (\Lya) absorbers.  
The balance of hydrogen photoionization and radiative recombination rates sets the opacity of the low-redshift 
IGM, which can be measured from the distribution of \Lya\ absorbers in redshift ($z$) and \HI\ column density, 
$N_{\rm HI}$ (cm$^{-2}$).  This allows one to infer the unidirectional ionizing photon flux 
$\Phi_0$ (cm$^{-2}$~s$^{-1}$)  and specific intensity $I_{\nu}$ (\uvunits) of  LyC radiation.  K14 found that 
the hydrogen photoionization rate ($\Gamma_{\rm H}$) required to match their simulated distribution of \HI\ 
absorbers was  {\it five} times larger than the value predicted in a recent calculation of the UVB by HM12.   
They found better agreement with previous estimates of a higher UVB (Shull \etal\ 1999; HM01) that included 
significant contributions from galaxies.   A large part of the UVB discrepancy can be traced to the anomalously 
low values at $z < 2$ of the LyC escape fraction, $f_{\rm esc} = (1.8 \times 10^{-4})(1+z)^{3.4}$, adopted by 
HM12 for galaxies.    
 
In this paper, we undertake a careful examination of this UVB discrepancy, comparing observations of 
low-$z$  \Lya\ absorber to new cosmological simulations\footnote{Our simulations are made 
with the N-body hydrodynamic grid code \texttt{Enzo} (http://enzo-project.org). For an overview of the code
see Bryan \etal\ (2014).  Our previous applications of \texttt{Enzo} to IGM astrophysics appear in 
Smith \etal\ (2011), Shull \etal\ (2012a), and Shull \etal\ (2012b).}.
Differences among previous theoretical estimates for the UVB may arise from modeling of the 
evolution of sources of EUV radiation from galaxies and quasars.   We use \HI\ data from our recent surveys 
of low-redshift \Lya\ absorbers with UV spectrographs on the {\it Hubble Space Telescope} (\HST):  
the STIS (Space Telescope Imaging Spectrograph) survey of 746 \Lya\ absorbers (Tilton \etal\ 2012) and 
the COS (Cosmic Origins Spectrograph) survey of over 2600 \Lya\ absorbers (Danforth \etal\ 2015, updated 
in July 2015).  These surveys give consistent results for the bivariate distribution, 
$f(N_{\rm HI}, z) \equiv d^2 {\cal N} / d (\log N_{\rm HI}) dz$, of absorbers in redshift ($z  \leq 0.47$) and 
\HI\ column density ($12.5 \leq \log N_{\rm HI} \leq 15.5$).


\begin{deluxetable}{lcc}
\tabletypesize{\footnotesize}
\tablecaption{\bf Hydrogen Photoionization Rates\tablenotemark{a} }  
\tablecolumns{4}
\tablewidth{0pt}
\tablehead{  \colhead {Model Reference} &  \colhead{$\Gamma_{\rm H} (z=0)$} &  \colhead{$\Gamma_{\rm H} (z=0.25)$}   } 
                                                                                     
\startdata
Haardt \& Madau (1996)            &   $4.14 \times 10^{-14}$   &  $8.63 \times 10^{-14}$   \\
Shull \etal\ (1999)                       &   $6.3 \times 10^{-14}$    &  $15 \times 10^{-14}$ \\
Haardt \& Madau (2001)            &   $10.3 \times 10^{-14}$   &  $20.6 \times 10^{-14}$   \\
Haardt \& Madau (2005)            &   $13.5 \times 10^{-14}$   &  $29.1 \times 10^{-14}$   \\
Haardt \& Madau (2012)            &   $2.28 \times 10^{-14}$   &  $5.89 \times 10^{-14}$   \\
Faucher-Gigu\`ere \etal\ (2009) &   $3.84 \times 10^{-14}$   &  $7.28 \times 10^{-14}$   \\
Shull \etal\ (2015)                       &   $4.6 \times 10^{-14}$     &  $12 \times 10^{-14}$ \\
\enddata  

\tablenotetext{a} {Hydrogen photoionization rates $\Gamma_{\rm H}$ (s$^{-1}$) computed at
redshifts $z = 0$ and $z = 0.25$ by various theoretical models.  \\ }

\end{deluxetable}


In Section 2, we describe current estimates of the UVB and low-redshift hydrogen photoionization rates and
compare our \HST\ surveys of intergalactic \HI\  column densities with new grid-code simulations of the IGM.  
We calibrate the UVB through its influence on the distribution of \HI\ column densities and the flux decrement 
($D_A$) from \Lya\ line blanketing.  In Section 3, we justify boosting $\Gamma_{\rm H}$ by a factor of 2--3 
above HM12 values, together with other parameters that characterize the UVB:  the specific intensity 
$I_0$ (\uvunits) at the Lyman limit and the integrated LyC photon flux $\Phi_0$ (cm$^{-2}$~s$^{-1}$).   
For $0 \leq z \leq 0.47$, our revised ionization rate is 
$\Gamma_{\rm H} \approx (4.6 \times 10^{-14}$ s$^{-1})(1+z)^{4.4}$.  At $z = 0$, this corresponds to  
$\Phi_0 \approx 5700 $ cm$^{-2}$~s$^{-1}$ and
$I_0 = 1.7 \times 10^{-23} \; {\rm erg~cm}^{-2}~{\rm s}^{-1}~{\rm Hz}^{-1}~{\rm sr}^{-1}$.
These parameters are in agreement with photoionization modeling in our recent IGM survey (Shull \etal\ 2014) 
of metal-ion ratios (\CIII/\CIV\ and \SiIII/\SiIV). This increase in the UVB is comparable to previous calculations 
(Shull \etal\ 1999; HM01, HM05) and likely arises from imprecise modeling of sources of EUV radiation (galaxies 
and quasars).  The larger UVB would be consistent with an increased contribution of galaxies with LyC escape 
fractions, $f_{\rm esc} \approx 0.05$ at $z < 2$, in contrast to the very low values $f_{\rm esc} < 10^{-3}$ adopted 
by HM12.  A recent analysis (Khaire \& Srianand 2015) of the effects of a revised QSO luminosity function 
(Croom \etal\ 2009; Palanque-Delabrouille \etal\ 2013) increased the low-redshift UVB by a factor of two.  Both 
of these results suggest that the UVB extrapolated to low redshift was under-estimated by HM12 by a factor of 
approximately 2--3.

 \section{CONSTRAINING THE UV BACKGROUND }
 
 \subsection{Definitions and Measurements of the Ionizing Radiation Field}  
  
 For an isotropic radiation field of specific intensity $I_{\nu}$, the normally incident {\it photon flux} per frequency 
 is $(\pi I_{\nu}/h \nu)$ into an angle-averaged, forward-directed effective solid angle of $\pi$ steradians.  The 
 isotropic photon flux striking an atom or ion is $4 \pi (I_{\nu} / h \nu)$, and the hydrogen photoionization rate 
 follows by integrating this photon flux times the photoionization cross section over frequency from threshold 
 ($\nu_0$) to $\infty$.  
\begin{equation} 
  \Gamma_{\rm H} =  \int _{\nu_0}^{\infty} \frac {4 \pi \, I_{\nu}}{h \nu} \sigma_{\nu} \, d \nu \approx
           \frac { 4 \pi I_0 \sigma_0} {h (\alpha+3) }   \; .
\end{equation}
Here, we approximate the frequency dependence of specific intensity and photoionization cross section by 
power laws, $ I_{\nu} = I_0  (\nu / \nu_0)^{-\alpha}$ and $\sigma_{\nu}  =  \sigma_0 (\nu / \nu_0)^{-3}$, where 
$\sigma_0 = 6.30 \times 10^{-18}$~cm$^2$ and $\alpha \approx 1.4$ for AGN (Shull \etal\ 2012c; Stevans 
\etal\ 2014).  The integrated unidirectional flux of ionizing photons is then
\begin{equation}
   \Phi_0 = \int_{\nu_0}^{\infty} \frac { \pi I_{\nu} } {h \nu} \; d \nu  = \frac {\pi I_0} {h \alpha} \;  \; ,
\end{equation} 
which is related to the density of hydrogen-ionizing photons by $\Phi_0 = n_{\gamma} (c/4)$ for an
 isotropic radiation field.  We then have the relations among parameters:  
\begin{eqnarray}
   \Gamma_{\rm H} &=& 4 \sigma_0 \Phi_0 \left(  \frac {\alpha} { \alpha + 3 } \right) = 
           (8.06 \times 10^{-14}~{\rm s}^{-1}) \Phi_4   \\
   \Gamma_{\rm H} &=&   \frac { 4 \pi I_0 \sigma_0 } { h ( \alpha + 3) }  = 
           (2.71\times10^{-14}~{\rm s}^{-1}) I_{-23}  \; \; ,  
\end{eqnarray}
where we normalize the incident flux of ionizing photons and specific intensity to characteristic values, 
$\Phi_0 = (10^4~{\rm cm}^{-2} \,  {\rm s}^{-1}) \Phi_4$ and 
$I_0 = (10^{-23} \; {\rm erg~cm}^{-2}~{\rm s}^{-1}~{\rm Hz}^{-1}~{\rm sr}^{-1}) I_{-23}$ at the hydrogen 
Lyman limit ($h \nu_0 = 13.60$~eV).   In an \HST/COS survey of IGM metallicity at $z \leq 0.4$,
Shull \etal\ (2014) found that $\Phi_4 \approx 1$ and $I_{-23} \approx 3$ gave reasonable fits to the observed 
ratios of adjacent ionization states of carbon and silicon, (\SiIII/\SiIV) $= 0.67^{+0.35}_{-0.19}$,  
(\CIII/\CIV) $= 0.70^{+0.43}_{-0.20}$, and their sum, 
$(\Omega_{\rm CIII} + \Omega_{\rm CIV}) / (\Omega_{\rm SiIII} + \Omega_{\rm SiIV}) = 4.9^{+2.2}_{-1.1}$.  
 
Over the past 20 years, numerous papers have estimated the ionizing background and photoionization rate.  
Table 1 lists $\Gamma_{\rm H}$ for several models, with values at $z = 0$ ranging from 
$(2.28-13.5) \times 10^{-14}$~s$^{-1}$.  The HM12 rate, $\Gamma_{\rm H} = 2.28 \times 10^{-14}~{\rm s}^{-1}$, 
corresponds to a one-sided ionizing flux 
$\Phi_0 = [\Gamma_{\rm H}  (\alpha + 3)/ 4 \sigma_0 \, \alpha] \approx 2630~{\rm cm}^{-2}~{\rm s}^{-1}$ and
specific intensity $I_{-23} \approx 0.823$ for the radio-quiet AGN spectral index ($\alpha = 1.57$) assumed by 
HM12.  These fluxes are lower by a factor of three compared to the $z = 0$ metagalactic radiation fields from 
AGN and galaxies calculated by Shull \etal\ (1999), $I_{\rm AGN} = 1.3^{+0.8}_{-0.5} \times 10^{-23}$ and
$I_{\rm Gal} = 1.1^{+1.5}_{-0.7} \times 10^{-23}$, respectively.   Adding these two values with propagated 
errors gives a total intensity and hydrogen ionization rate of
$I_{\rm tot} =  2.4^{+1.7}_{-0.9} \times 10^{-23} \; {\rm erg~cm}^{-2}~{\rm s}^{-1}~{\rm Hz}^{-1}~{\rm sr}^{-1}$ and
$\Gamma_{\rm H}  =  6.0^{+4.2}_{-2.1} \times 10^{-14} \; {\rm s}^{-1}$.  The difference between these radiation 
fields appears to be the contribution from galaxies.   The HM12 background adopts a negligible UVB from galaxies,
whereas the background models of Shull \etal\ (1999), HM01, and HM05 have comparable intensities from
galaxies, owing to higher assumed LyC escape fractions.  
  
 \subsection{\HST\ Observations of the \HI\ Column Density Distribution}
  
Previous ultraviolet spectroscopic surveys of low-redshift \Lya\ absorbers estimated their contribution to the baryon 
census through the distribution of \HI\ column densities in the diffuse \Lya\ forest (e.g., Penton \etal\ 2000, 2004; 
Lehner \etal\ 2007; Danforth \& Shull 2008).  Our recent \HST\ surveys of the low-redshift IGM were more extensive, 
obtaining 746 \Lya\ absorbers with STIS (Tilton \etal\ 2012) and 2577  \Lya\ absorbers with COS (Danforth \etal\ 2015).   
The COS survey used the medium-resolution far-UV gratings (Green \etal\ 2012) with coverage between 
1135--1460~\AA\ (G130M) and 1390--1795~\AA\ (G160M); a few spectra extended slightly outside these boundaries.  
Our COS survey probed 82 AGN sight lines with cumulative pathlength $\Delta z = 21.7$ with \HI\ column densities 
$N_{\rm HI}$ (in cm$^{-2}$) between $12.5 <  \log N_{\rm HI} < 17$.  For this paper, we analyze a ``uniform 
redshift-limited sample" of \Lya\ absorbers at $z \leq 0.47$.  As discussed in these survey papers, the \Lya\ statistics 
are consistent with 24--30\% of the baryons residing in the \Lya\ forest and partial Lyman-limit systems.


\begin{deluxetable}{ccccc}
\tabletypesize{\footnotesize}
\tablecaption{\bf Column Density Distribution\tablenotemark{a}(STIS Survey) }  
\tablecolumns{5}
\tablewidth{0pt}
\tablehead{  \colhead { $\langle \log N_{\rm HI} \rangle$} &  \colhead{Range in $\log N_{\rm HI}$} &
     \colhead{ $\cal{N}_{\rm abs}$ } &  \colhead{ $\Delta z_{\rm eff}$ } &  \colhead {$f(N_{\rm HI}, z)$ } 
  }
                                                                                     
\startdata
13.0   &   (12.9--13.1)  &  115   &  4.637  &  $124^{+13}_{-13}$    \\
13.2   &   (13.1--13.3)  &  115   &  5.003  &  $115^{+11}_{-11}$    \\
13.4   &   (13.3--13.5)  &  106   &  5.216  &  $102^{+10}_{-10}$   \\
13.6   &   (13.5--13.7)  &    75   &  5.327  &  $70^{+9}_{-8}$         \\
13.8   &   (13.7--13.9)  &    73   &  5.341  &  $68^{+9}_{-8}$         \\
14.0   &   (13.9--14.1)  &    50   &  5.360  &  $47^{+8}_{-7}$         \\
14.2   &   (14.1--14.3)  &    35   &  5.379  &  $33^{+6}_{-5}$        \\
14.4   &   (14.3--14.5)  &    26   &  5.382  &  $24^{+6}_{-5}$        \\
14.6   &   (14.5--14.7)  &    18   &  5.382  &  $17^{+5}_{-4}$        \\
\enddata

\tablenotetext{a} {For 613 low-redshift \Lya\ absorbers taken from \HST/STIS survey (Tilton \etal\ 2012),
   the columns show:  (1) mean column density ($N_{\rm HI}$ in cm$^{-2})$
   with bin width $\Delta \log N_{\rm HI} = 0.2$; (2)  bin range in $\log N_{\rm HI}$; (3) number of \Lya\ 
   absorbers ($\cal{N}_{\rm abs}$) in bin;  (4) total redshift pathlength $\Delta z_{\rm eff}$ 
   at each $N_{\rm HI}$; (4) bivariate distribution of absorbers in column density and redshift, 
  $f(N_{\rm HI}, z) \equiv d^2 {\cal N} / d(\log N_{\rm HI}) dz$, computed as 
  $({\cal N}_{\rm abs} / 0.2 \, \Delta z_{\rm eff})$ and plotted in Figures 2--7. \\  }

\end{deluxetable}



\begin{deluxetable}{ccccc}
\tabletypesize{\footnotesize}
\tablecaption{\bf Column Density Distribution\tablenotemark{a} (COS Survey) }  
\tablecolumns{5}
\tablewidth{0pt}
\tablehead{  \colhead { $\langle \log N_{\rm HI} \rangle$ } &  \colhead{Range in $\log N_{\rm HI}$} &
     \colhead{ $\cal{N}_{\rm abs}$ } & \colhead{ $\Delta z_{\rm eff}$ } &  \colhead { $f(N_{\rm HI}, z)$} 
  }
                                                                                     
\startdata
12.7   &   (12.6--12.8)  &  194  &     3.67     &  $260^{+1000}_{-120}$    \\
12.9   &   (12.8--13.0)  &  292   &    9.79     &  $150^{+80}_{-40}$   \\
13.1   &   (13.0--13.2)  &  383   &  16.46     &  $120 \pm 10$    \\
13.3   &   (13.2--13.4)  &  368   &  18.78     &  $98 \pm 5$   \\
13.5   &   (13.4--13.6)  &  267   &  19.23     &  $69^{+17}_{-12}$    \\
13.7   &   (13.6--13.8)  &  146   &  12.03     &  $61 \pm 5$    \\
13.9   &   (13.8--14.0)  &  123   &  12.25     &  $50 \pm 5$    \\
14.1   &   (14.0--14.2)  &    96   &  12.96     &  $37 \pm 4$    \\
14.3   &   (14.2--14.4)  &    75   &  13.75     &  $27^{+4}_{-3}$    \\
14.5   &   (14.4--14.6)  &    47   &  14.17     &  $17^{+3}_{-2}$    \\
14.7   &  (14.6--14.8)   &    45   &  14.24     &  $16^{+3}_{-2}$    \\
14.9   &   (14.8--15.0)  &    19   &  14.25     &  $6.7^{+1.9}_{-1.5}$    \\
15.1   &   (15.0--15.2)  &    19   &  14.25     &  $6.7^{+1.9}_{-1.5}$    \\
\enddata

\tablenotetext{a} {Statistics of low-redshift \Lya\ absorbers ($12.6 \leq \log N_{\rm HI} < 15.2$) 
   taken from the \HST/COS survey (Danforth \etal\ 2015, updated July 2015) using the 
   ``uniform sub-sample" of 2074 \HI\ absorbers. The columns show:  
   (1) mean column density ($N_{\rm HI}$ in cm$^{-2}$) with bin width $\Delta \log N_{\rm HI} = 0.2$;
   (2)  bin range in $\log N_{\rm HI}$; (3) number of \Lya\ absorbers ($\cal{N}_{\rm abs}$) in bin; 
   (4) total redshift pathlength $\Delta z_{\rm eff}$ at each $N_{\rm HI}$; (4) bivariate distribution of 
   absorbers in column density and redshift, 
  $f(N_{\rm HI}, z) \equiv d^2 {\cal N} / d(\log N_{\rm HI}) dz$, computed as  
   $({\cal N}_{\rm abs} / 0.2 \, \Delta z_{\rm eff})$ and plotted in Figures 2--7. \\ }

\end{deluxetable}


Table 2 summarizes the data selected from the STIS survey, with 613 \Lya\ absorbers at $z\leq 0.4$ over the range 
$12.9 \leq \log N_{\rm HI} \leq 14.7$, reprocessed as described in Danforth \etal\ (2015). Table 3 gives similar results 
from the COS survey at $z \leq 0.47$, with 2074 absorbers between $12.6 \leq \log N_{\rm HI} \leq 15.2$.  For each 
bin in $\log N_{\rm HI}$, we list the number of absorbers ($\cal{N}_{\rm abs}$) and effective redshift pathlength
($\Delta z_{\rm eff}$) over which our \Lya\ survey is sensitive.  From these data, we derive the bivariate distribution, 
$d^2 {\cal N} / d(\log N_{\rm HI}) dz$, of absorbers in \HI\ column density and redshift by dividing $\cal{N}_{\rm abs}$ 
by $\Delta z_{\rm eff}$ and $\Delta \log N_{\rm HI} = 0.2$.  An accurate calculation of this distribution function
requires both good absorber counting statistics and knowledge of the pathlength $\Delta z_{\rm eff}$ covered in the 
survey for a given column density.  As in Danforth \& Shull (2008), we compute $\Delta z_{\rm eff}$  corresponding to 
the $4 \sigma$ minimum \Lya\ equivalent width as a function of wavelength in each spectrum.   Asymmetric error bars 
arise from statistical uncertainties in $N_{\rm abs}$ and $\Delta z_{\rm eff}$, computed for each bin using the formalism 
of Gehrels (1986).  At low column densities, the errors are dominated by $\Delta z_{\rm eff}$, while at high column 
densities the errors are dominated by the small values of $N_{\rm abs}$ in bins at $\log N_{\rm HI} \geq 14.6$.

An important effect of the IGM is the flux decrement, $D_A$, produced by \Lya\ forest line blanketing of the continuum flux.   
Previous measurements of $D_A$ were reported by Kirkman \etal\ (2007) using low-resolution UV observations with the
\HST\ Faint Object Spectrograph (FOS) toward 74 AGN.  Their measured decrements were fitted to a power law,  
$D_A(z) = (0.016)(1+z)^{1.01}$, over the range $0 < z < 1.6$ with considerable scatter.   Based on our moderate-resolution
HST/COS survey,  Figure~1 shows the fraction of light removed by \Lya\ absorption from the continuum for $0 < z < 0.47$.
These statistics were found by summing the {\it observed-frame} equivalent widths of all identified ($>4\sigma$) \Lya\ 
absorbers in the Danforth \etal\ (2015) catalog within a given redshift range, normalized by the effective redshift
pathlength probed by the survey for absorbers of that strength.  The uncertainty, $\sigma_{D_A}$, is the quadratic sum 
of measured equivalent-width uncertainties normalized in the same manner.  The errors are small at lower redshifts 
($z \la 0.35$) where most of the surveyed \Lya\ absorbers are found.  However, they remain small compared to $D_A$
even at higher redshifts where $\Delta z_{\rm eff}$ is smaller.   Figure~1 shows our results for bins of width 
$\Delta z = 0.01$ and $\Delta z=0.05$ to illustrate the variance on small scales.  The scatter in $D_A$ is significantly 
larger than $\sigma_{D_A}$, and cosmic variance is significant, particularly at $z > 0.35$.   These fluctuations are 
large in the $\Delta z = 0.01$ bins, which are comparable to the mean absorber separations for the observed  \Lya\ 
absorption-line frequencies $d {\cal N}/dz \approx 100-200$ for $\log N_{\rm HI} = 12.7-13.3$.  

Figure 1 shows a small dip in $D_A$ between $0.2 < z < 0.3$.  This redshift band corresponds to \Lya\ absorbers at 
1459~\AA\ to 1580~\AA, observed primarily by the COS/G160M grating (1390--1795~\AA) since G130M ends at 
1460~\AA.   The dip appears in several ranges of column density, $13 < \log N_{\rm HI} < 14$ and 
$14 < \log N_{\rm HI} < 15$.  The first range is expected to dominate line blanketing in the \Lya\ forest for line profiles 
with Doppler velocity parameters $b \approx 25-30$~\kms.  The dip also remains when we offset the redshift bins. Thus, 
we believe the dip to be real, although we do not have a plausible physical explanation for its presence.  

We fitted the mean values of the \HST\ observations with $\Delta z = 0.05$ bins to the power-law form
$D_A(z) = (0.014 \pm 0.001)(1+z)^{2.2 \pm 0.2}$.  For $z \leq 0.2$, these observations are in reasonable agreement 
with the lower-resolution \HST/FOS results of Kirkman \etal\ (2007), although we find steeper redshift evolution. 
By post-processing the \HI\ absorption-line profiles in our simulations (Sections 2.3 and 2.4) over the range $0 < z < 0.2$, 
we find mean decrements at $\langle z \rangle = 0.1$ of $D_A = 0.0116$, 0.0097, and 0.0332 for the HM01, HM05, and 
HM12 radiation fields, respectively.   At  $\langle z \rangle = 0.1$, these UVBs correspond to hydrogen photoionization rates 
$\Gamma_{\rm H} = 3.54 \times 10^{-14}~{\rm s}^{-1}$ (HM12),  $1.38 \times 10^{-13}~{\rm s}^{-1}$ (HM01), and 
$1.86 \times 10^{-13}~{\rm s}^{-1}$ (HM05).   
Our simulations find that $D_A \propto \Gamma_{\rm H}^{-0.7}$.  In their SPH simulations, K14 found 
decrements at $z \approx 0.1$ of $D_A = 0.024$ (HM01) and $D_A = 0.050$ (HM12), higher than our simulations by
factors of 1.9 and 2.9, respectively.  Similar offsets are seen in the distribution of \HI\ column densities.


\begin{figure}
  \epsscale{1.2}
  \plotone{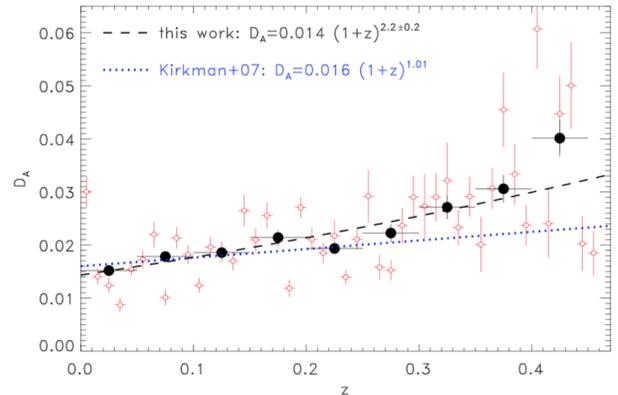}
\caption{\small{Flux decrement from the Danforth \etal\ (2015) low-$z$ IGM survey with COS (updated
in July 2015) showing the 
fraction of flux removed from the continuum by \Lya\ absorbers for redshifts $0 < z < 0.47$. The decrement 
$D_A(z)$ is calculated from the normalized, summed,observed-frame equivalent widths of \Lya\ absorbers 
in a given redshift bin, divided by the effective pathlength ($\Delta z_{\rm eff}$) probed by the survey at each 
redshift.  Red open circles show $D_A$ in bins of width $\Delta z = 0.01$.  Filled black circles show $D_A$ 
in bins of $\Delta z = 0.05$ and fitted to $D_A(z) = (0.014)(1+z)^{2.2 \pm 0.2}$  (dashed line).
Blue dotted line shows the fit, $D_A(z) \approx (0.016)(1+z)^{1.01}$, found by  Kirkman \etal\ (2007) in 
lower-resolution \HST/FOS measurements.  The small drop in $D_A$ at $z \approx 0.2-0.3$ appears over 
several column-density ranges and remains when we shift redshift bins. \\  }
 }
\end{figure}


Our observations at $z \leq 0.47$ with STIS (Tilton \etal\ 2012) and COS (Danforth \etal\ 2015) yield consistent values 
for  $f(N_{\rm HI}, z)$, the bivariate distribution of \Lya\ absorbers in column density and redshift. A least-squares fit of 
the COS data over the range  $12.7 \leq \log N_{\rm HI} \leq 15.2$ had the form,
$f(N_{\rm HI}, z) \equiv d^2 {\cal N} / d(\log N_{\rm HI}) dz \approx (167) N_{13}^{-0.65\pm0.02}$, 
where we define the dimensionless column density $N_{13} = [N_{\rm HI}/10^{13}~{\rm cm}^{-2}]$.  We can
use this line frequency to make an analytic estimate of the \Lya\ line blanketing and flux decrement,
\begin{eqnarray}
   D_A  &=& \int_{N_1}^{N_2}  \left( \frac {W_{\lambda}} {\lambda} \right) f(N,z) \, d(\log N)  \nonumber   \\
   &=& \left( \frac {\pi e^2 f \lambda} {m_e c^2} \right) \left( \frac {167 \times 10^{13}~{\rm cm}^{-2} } {2.303} \right)
    \int_{N_1}^{N_2} N_{13}^{-0.65} \, dN_{13}   \; .
\end{eqnarray} 
In the last expression of Equation (5), we assume unsaturated \Lya\ lines, with equivalent widths
$W_{\lambda}/\lambda = (\pi e^2/m_e c)(Nf\lambda/c)$, where $W_{\lambda}$ and $\lambda = 1215.67$~\AA\ 
are defined in the rest-frame and $f = 0.4164$ is the \Lya\ oscillator strength.  We adopt limits $\log N_1 = 12.5$ 
and $\log N_2 = 14.0$ for the integral, which we denote $I(N_1, N_2) \approx 4.49$.  The estimated decrement 
$D_A = (0.00323) I(N_1, N_2) \approx 0.0145$, close to our observations at $z \approx 0$.  Because line frequency 
$f(N_{\rm HI}, z) \propto (1+z)^{2.2}$ out to $z \approx 0.5$ (Danforth \etal\ 2015), we expect $D_A$ to have the same 
dependence, as seen in Figure~1.  This calculation shows that for the observed steep distribution of \HI\ column 
densities, most of the line blanketing occurs from moderate-strength \Lya\ absorbers at $N_{\rm HI} \approx N_2$, 
beginning to saturate and appear on the flat portion of the curve of growth.   Owing to line saturation, the stronger 
but rarer \Lya\ absorbers add a small amount to this estimate. 
  
\subsection{Cosmological Simulations of  \HI\ in the low-$z$ IGM}  
    
Our simulations of the \ion{H}{1} absorber distributions were run using the Eulerian N-body + hydrodynamics code 
\texttt{Enzo} (Bryan \etal\ 2014). The N-body dynamics of the dark matter particles in our simulations were calculated 
using a particle-mesh solver (Hockney \& Eastwood 1988), and the hydrodynamic equations were solved using a 
direct-Eulerian piecewise parabolic method (Colella \& Woodward 1984; Bryan \etal\ 1995) with a 
Harten-Lax-van Leer-Contact Riemann solver (Toro \etal\ 1994).  Our modifications of this code and its applications 
to the IGM are discussed in Smith \etal\ (2011), where we conducted tests of convergence, examined various feedback 
schemes, and compared the results to IGM  thermal phases, \OVI\ absorbers, and star-formation histories using 
box sizes of $25 h^{-1}$~Mpc and $50 h^{-1}$~Mpc and grids of $256^3$, $384^3$, $512^3$, $768^3$, and $1024^3$ 
cells.  We note that the Smith \etal\ (2011) simulations used the HM96 background, rather than HM01 as stated in that 
paper\footnote{Because the HM96 background is 2.5 times lower than HM01, the \Lya\ forest was stronger in the Smith 
\etal\ (2011) simulations.  However, as we discuss in Section 2.4, the HM96 background is close to the value we 
recommend in our current study.   Consequently, our 2011 simulations are in reasonable agreement with the \HST\ 
observations of the low-redshift \Lya\ forest.}.
This code has a substantial IGM heritage, with applications to \OVI\ absorbers (Smith \etal\ 2011), the baryon census 
at $z < 0.4$ (Shull \etal\ 2012b), clumping factors and critical star-formation rates during reionization (Shull \etal\ 2012a), 
and synthetic absorption-line spectra for comparison of simulations with \HST\ observations  (Egan \etal\ 2014).  

Our  current simulations were initialized at a redshift $z = 99$ and run to $z = 0$, using the WMAP-9 maximum
likelihood concordance values (Hinshaw \etal\ 2013) with $\Omega_m = 0.282$, $\Omega_\Lambda = 0.718$, 
$\Omega_b = 0.046$, $H_0 = 69.7$~km~s$^{-1}$~Mpc$^{-1}$, $\sigma_8 = 0.817$, and $n_s = 0.965$
to create the initial conditions.  The dark matter density power spectrum used the Eisenstein \& Hu (1999) transfer 
function.  Three independent realizations of the initial conditions were created in order to constrain the effects of cosmic 
variance due to the finite box size. Two of the realizations were created using \textsc{Music} (Hahn \& Abel 2011) with 
second-order Lagrangian perturbation theory. The third realization was created using the initial condition generator packaged 
with \texttt{Enzo}.  The basic simulations discussed in this paper were run on static, uniform grids with a box size of $50h^{-1}$ 
comoving Mpc and $512^3$ and $768^3$ cells.   We also analyzed a previous simulation with $1536^3$ cells, produced by 
Britton Smith on the XSEDE supercomputer and discussed in our study of IGM clumping factors (Shull \etal\ 2012a), as well
the ($1024^3, 50 h^{-1}$~Mpc) simulations of Smith \etal\ (2011).  Because those $1024^3$ models were run with an 
older UVB from Haardt \& Madau (1996), we chose not to compare them to our $512^3$ and $768^3$  runs, which used 
more recent backgrounds (HM01, HM05, HM12).  

In view of the different predictions of our code for the distribution of low-redshift \Lya\ absorbers, it is appropriate to 
compare the resolution\footnote{The original study by Dav\'e \etal\ (2010) used simulations in a $48 h^{-1}$~Mpc box, 
and most of their studies were run with $384^3$ particles.  To investigate effects of numerical resolution and box size, 
they also employed a simulation with $512^3$ particles in a $96 h^{-1}$~Mpc box.  Contemporaneous SPH studies 
(Tepper-Garcia \etal\ 2012) of the IGM were conducted by the ``OverWhelmingly Large Simulations" (\textsc{OWLS})
project (Schaye \etal\ 2010) in a  $100h^{-1}$~Mpc box with $512^3$ dark-matter (and baryon) particles.}
of our grid-code simulations to the SPH simulations run by K14.  With a co-moving box size of $50h^{-1}$~Mpc, our 
models have spatial resolutions (cell sizes) of $97.6h^{-1}$, $65.1h^{-1}$, and $32.6h^{-1}$~kpc 
for grids of $512^3$, $768^3$, and $1536^3$, respectively.  The corresponding dark-matter mass resolutions are 
$m_{\rm dm} = 6.1 \times 10^7 \, h^{-1}$~\msun\ ($512^3$),  $1.8 \times 10^7 \, h^{-1}$~\msun\ ($768^3$), and 
$2.2 \times 10^6 \, h^{-1}$~\msun\ ($1536^3$).  For comparison, the simulations analyzed by K14 used the SPH code 
of Dav\'e \etal\ (2010), also in a $50 h^{-1}$~Mpc  box with $576^3$ dark-matter (and baryon) particles.  The characteristic 
spatial resolution in their SPH method is the interparticle distance, $50 h^{-1}\, {\rm Mpc} / 576 \approx 86.8 h^{-1}$~kpc.  
Under gravitational evolution, the SPH particles are concentrated in regions with baryon overdensity
$\Delta_b = \rho_b/ \bar{\rho}_b > 1$, and the interparticle distance is reduced by a factor of $\Delta_b^{1/3}$.  
Although K14 do not quote a mass resolution, their simulation ($50 h^{-1}$~Mpc, $576^3$) would have
$m_{\rm dm} \approx 5 \times10^7 h^{-1}$  \msun, scaling from values quoted in the ($512^3$, $100h^{-1}$ Mpc) SPH 
calculation of Tepper-Garcia \etal\ (2012).   Therefore, the mass resolution in our $512^3$ runs is comparable to that of K14, 
and our $768^3$ and $1536^3$ resolutions are superior.  Resolution can be important when applied to the large 
(100-200 kpc) \Lya\ absorption systems observed in the low-redshift IGM.  

Star particles were formed in the simulations using the subgrid physics prescription of Cen \& Ostriker (1992). Star formation 
occurs in cells where the baryon density is greater than 100 times the critical density, the divergence of the velocity is negative, 
and the cooling time is less than the dynamical time.   As described by Smith \etal\ (2011), we adopted a star formation efficiency 
of 10\% in these cells.  Feedback from star formation returns gas, matter, and energy from star particles to the ISM and 
IGM.   Although star particles are formed instantaneously, feedback occurs gradually with an exponential decay over time. 
Star particles produce 90\% of their feedback within four dynamical times of their creation. Of the star particles' total mass, 
25\% is returned as gas, with 10\% of that taking the form of metals and $10^{-5}$ of the rest-mass energy returned as thermal 
feedback. The feedback is distributed evenly over 27 grid cells centered on the star particle. This scheme avoids the overcooling 
and subsequent runaway star formation that occurs when feedback is returned to a single cell.  We analyzed this ``distributed 
feedback" method in our previous study (Smith \etal\ 2011) and found that it produces better convergence and removes the 
over-cooling.   These simulations were able to reproduce both the observed star-formation history and the number density of 
\OVI\ absorbers per unit redshift over the range $0 < z < 0.4$.  Many SPH studies of the IGM (Dav\'e \etal\ 2010; Tepper-Garcia 
\etal\ 2012) suppress the over-cooling by suspending the cooling or turning off the hydrodynamics for a period of time, allowing 
the injected energy and metals to diffuse into the surrounding gas.  

The ionization states of H and He in the IGM were calculated with the non-equilibrium chemistry module in \texttt{Enzo} 
(Abel \etal\ 1997;  Anninos \etal\ 1997).   Metal cooling of the gas was computed using precompiled 
\textsc{Cloudy}\footnote{http://nublado.org} (Ferland 2013) tables that assume ionization equilibrium for the metals and are 
coupled to the chemistry solver (Smith \etal\ 2008). Photoionization and radiative heating of the gas come from a spatially uniform 
metagalactic UVB. For each realization of the initial conditions, we ran three simulations that were identical except for the form 
of the UVB.   Two simulations used analytic fits to the HM01 and HM05 backgrounds, initialized at $z = 8.9$ and reaching full 
strength by $z = 8$.  A third background used the HM12 table implemented in the 
\texttt{Grackle}\footnote{https://grackle.readthedocs.org/} chemistry and cooling library  (Bryan \etal\ 2014; Kim \etal\ 2014) that
begins at $z = 15.13$.   

In order to analyze the distribution of \Lya\ absorbers in the simulations, we created synthetic quasar sight lines using the 
YT\footnote{http://yt-project.org} analysis package (Turk \etal\ 2011).  For each simulation box, we created 200 synthetic sight 
lines, each constructed from simulation outputs spanning the redshift range $0.0 \leq z \leq 0.4$. A randomly oriented ray was 
chosen through each output, representing the portion of the sight line beginning at the output redshift and with sufficient pathlength 
to cover the $\Delta z$ to the next output file. These rays were then combined to create a sight line spanning the entire redshift range.
We used identical random seeds for simulations sharing the same initial conditions; differences in the absorber distributions 
therefore directly reflect effects of different ultraviolet backgrounds. Each sight line is composed of many line elements, from the 
portions of the sight line passing through a single grid cell.  We identified \Lya\ absorbers as sets of contiguous line elements with 
neutral hydrogen density $n_{\rm HI} \geq 10^{-12.5}~{\rm cm}^{-3}$.  Egan \etal\ (2014) compared this method of identifying 
absorbers to a more sophisticated method involving synthetic spectra.  In their Figure 11, they showed that the resulting absorber 
column density distributions are in good agreement.   We also analyzed a large ($1536^3, 50 h^{-1}$ Mpc) simulation run by 
Britton Smith and discussed in Shull \etal\ (2012a).  The light rays from this simulation were generated by Devin Silvia, and we 
analyzed them in the same manner.  

\subsection{Comparison of Simulations with Observations} 

The flux decrement $D_A$ is an imperfect metric for estimating the UVB, since measurements of the 
observed flux decrement depend on spectral resolution.  In addition, both measurements and simulations exhibit significant 
variance in $D_A$.  A more robust IGM diagnostic comes from the distribution of \HI\ column densities.  Figure 2 illustrates the 
observed STIS and COS distributions, $f(N_{\rm HI}, z)$, over the range $12.5 \leq \log N_{\rm HI} \leq 14.5$.  The STIS and 
COS bins are offset by $\Delta \log N_{\rm HI} = 0.1$, using data from Tables 2 and 3.  The COS survey is more extensive in 
both absorber numbers and column density, and its statistical accuracy is superior to that of the STIS survey.  Over the 
overlapping range in $\log N_{\rm HI}$,  the agreement is good.  We over-plot values of $f(N_{\rm HI}, z)$ from our new
IGM simulations ($50 h^{-1}$ Mpc box and $768^3$ grid cells) using three different ionizing backgrounds (HM01, HM05, HM12).  
As analyzed further in Section 2.5, the inferred baryon density requires ionization corrections for the neutral fraction, 
$n_{\rm HI} / n_H$.  Appendix A suggests that the amplitude of the \Lya\ absorber distribution, $f(N_{\rm HI}, z)$, is proportional 
to $\Gamma_{\rm H}^{-1/2} T_4^{-0.363}$ for fixed baryon content ($\Omega_b$) and temperature $T = (10^4~{\rm K})T_4$, 
and it therefore provides a constraint on the UVB. In our simulations, we find that 
$f(N_{\rm HI},z) \propto \Gamma_H^{-0.773}$ for absorbers in the range $13 < \log N_{\rm HI} < 14$.



\begin{figure}
 \epsscale{1.25}
\plotone{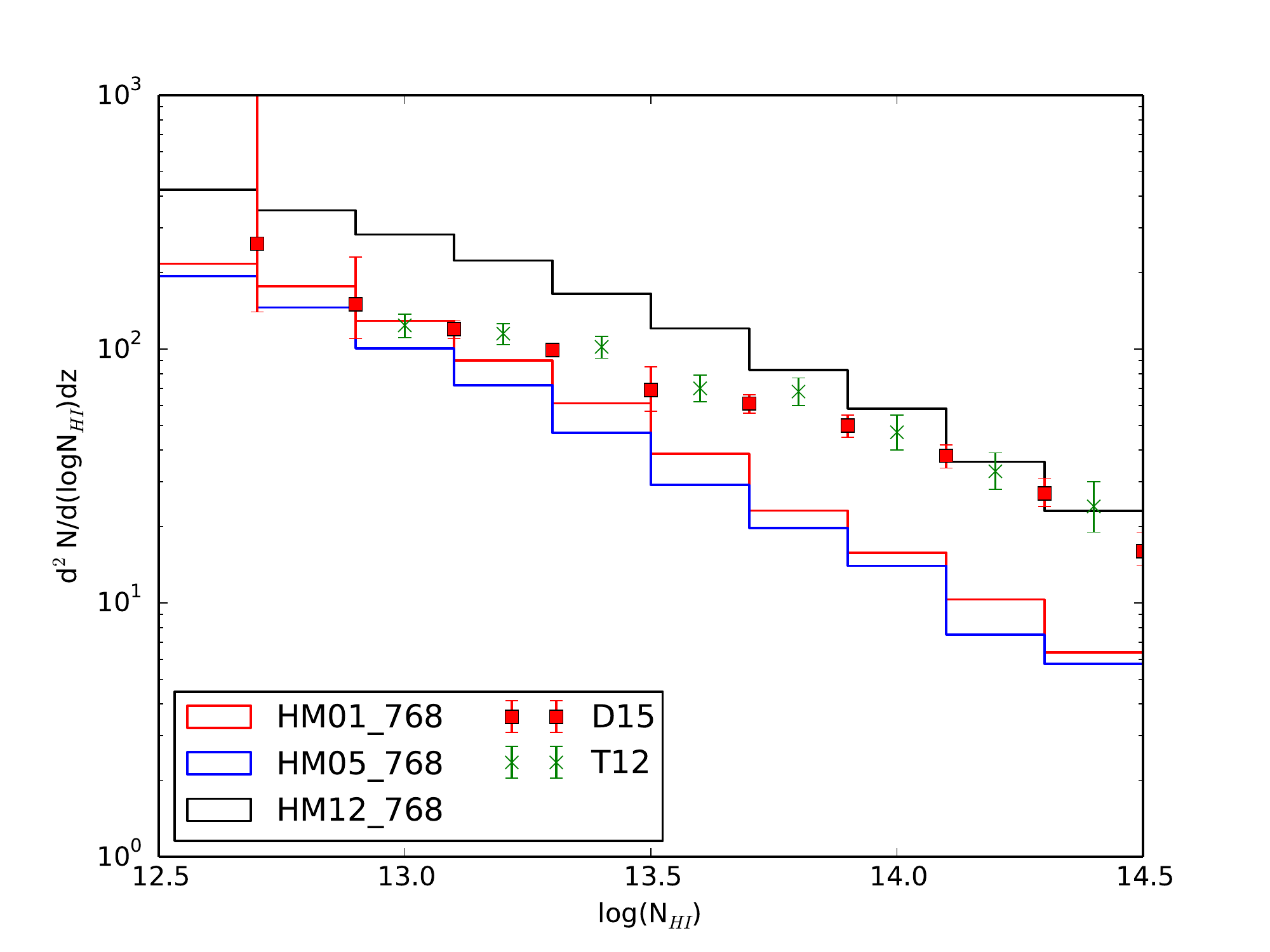}
\caption{ \small {\HST\ surveys of the column-density distribution of low-redshift IGM absorbers are compared to 
new cosmological simulations with the \texttt{Enzo} grid code ($768^3$ cells, $50h^{-1}$ Mpc box) with three
ionizing UV backgrounds (HM01, HM05, HM12).  We plot the bivariate distribution of absorbers, 
$f(N_{\rm HI}, z) = d^2 {\cal N} / d (\log N_{\rm HI} ) dz$, in redshift and column density ($N_{\rm HI}$ in cm$^{-2}$) 
with $\Delta \log N_{\rm HI} = 0.2$ bins.  Data from \HST/COS survey (D15: Danforth \etal\ 2015) are shown as red 
squares, and data from \HST/STIS survey (T12: Tilton \etal\ 2012) as green crosses.  Analytic theory (Appendix~A) 
shows that, for fixed baryon content, \Lya\ line frequency scales with photoionization rate and gas temperature as 
$\Gamma_{\rm H}^{-1/2} T_4^{-0.363}$.  The HM12 simulations use a lower ionizing flux and produce a higher 
absorption-line density.  For data with the best statistics ($13 \leq \log N_{\rm HI} \leq 14$), the observed distribution,
$f(N_{\rm HI}, z) \approx (167) [N_{\rm HI}/10^{13}~{\rm cm}^{-2}]^{-0.65}$, lies between simulations run
with higher UVB (HM01 and HM05) and the lower HM12 background, except at the highest column densities. 
The best-fit distribution requires an ionizing background approximately 2-3 times higher than HM12, but with 
sufficient uncertainty that a crisis in photon production is unlikely.} \\ } 

\end{figure}


Figures 2--7 illustrate the differential absorber distribution, $d^2N/d\left(\log N_{\rm HI}\right)dz$, with tests of 
convergence, cosmic variance, box size, feedback, and redshift evolution.  The simulated distributions are compared to 
observed distributions from \HST\ surveys of Tilton \etal\ (2012) and Danforth \etal\ (2015).  Figure~2 presents the 
primary results of this paper:  the \HI\ column-density distributions for three radiation fields (HM01, HM05, HM12) 
computed with $768^3$ grids and a $50 h^{-1}$~Mpc box. The lower HM12 background produces significantly more 
absorbers than earlier UVBs, ranging from a factor of two for absorbers between $12.5 \leq \log N_{\rm HI}  \leq 12.7$ to 
a factor of seven for $14.3 \leq \log N_{\rm HI} \leq 14.5$.  The observed distributions fall between the results for the HM01 
and HM12 backgrounds, approaching the HM12 simulations for $\log N_{\rm HI} > 14.0$. This suggests that there 
may be less tension between the HM12 background and observations than implied by the K14 results.  Figure 3 explores 
effects of cosmic variance, comparing three models with $50 h^{-1}$~Mpc boxes and $512^3$ grids.  For each 
radiation field (HM01, HM05, HM12) the simulations labeled with subscripts 1 and 2 are those with the \textsc{Music} 
initial conditions, and those labeled 3 use the \texttt{Enzo}-packaged initial conditions.  For a fixed UVB the scatter 
among simulations is at the $15\%$ level or less,  indicating that variance is unlikely to play an important role. 
Figure 4 explores convergence of our simulations, comparing models with $50 h^{-1}$~Mpc boxes on grids
of $512^3$, $768^3$, and $1536^3$.  The differences between these models are relatively small, at the 10\% level
and within the expected differences arising from sample variance.  Figure 5 investigates the potential effects of box size,
which could suppress structure formation down to low redshift if the box was too small.  We ran two simulations 
($100 h^{-1}$~Mpc, $512^3$ and $50 h^{-1}$~Mpc, $256^3$), both using the same HM01 ionizing background.  
The grid cell sizes are identical ($195 h^{-1}$~kpc), and the results are essentially the same for $\log N_{\rm HI} < 14$.



\begin{figure}
\epsscale{1.2}
 \plotone{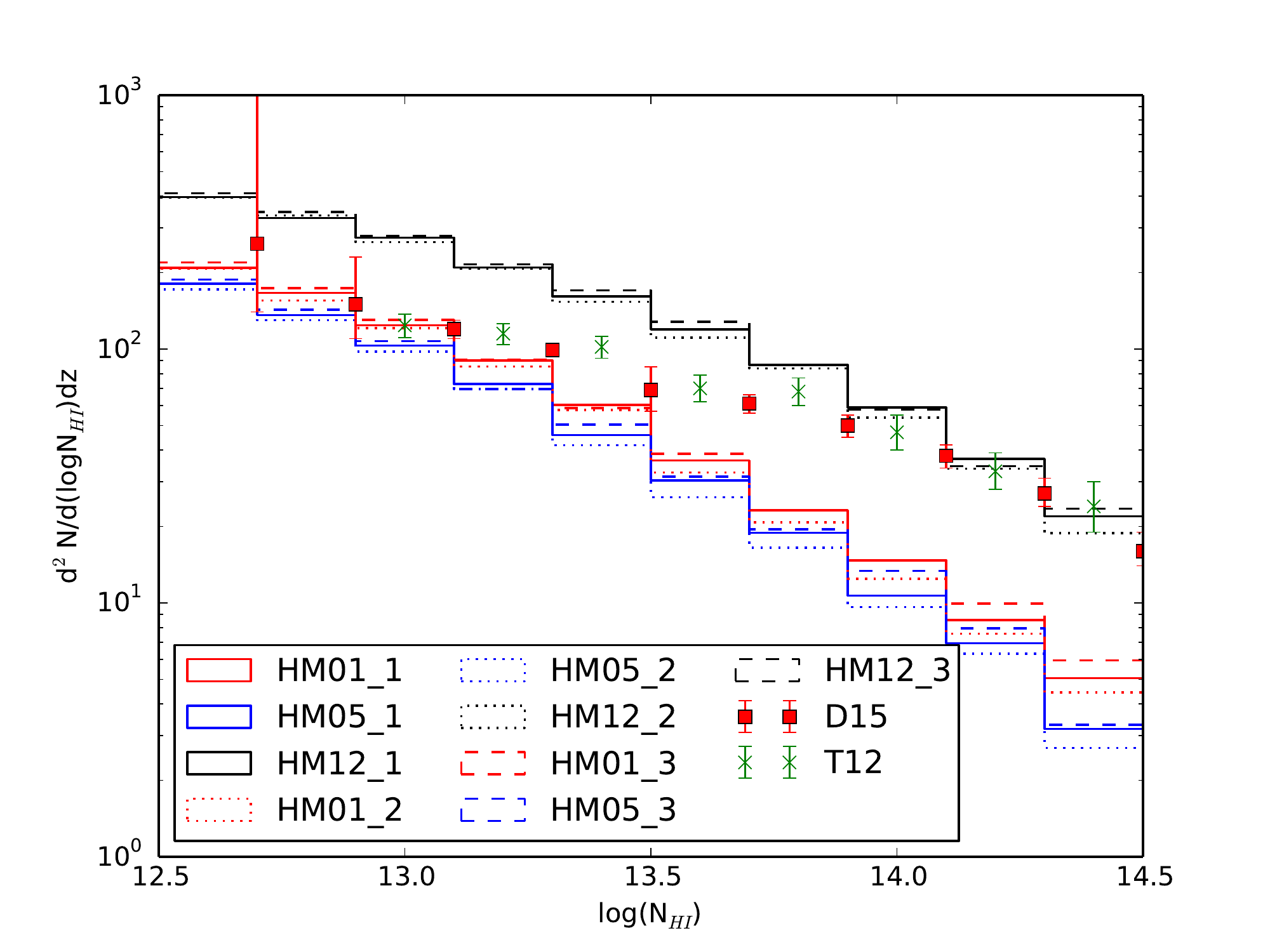}
\caption{ \small{Same format as Figure 2.  To test cosmic variance, we show three
($50 h^{-1}$~Mpc, $512^3$) simulations for each UVB, labeled HM01$\_i$, HM05$\_i$, 
and HM12$\_i$ (subscripts $i = 1, 2, 3$ correspond to runs with different initial conditions) 
and compared to \HST\ survey data (D15 and T12). \\ }
} 
\end{figure}




\begin{figure}
 \epsscale{1.2}
 \plotone{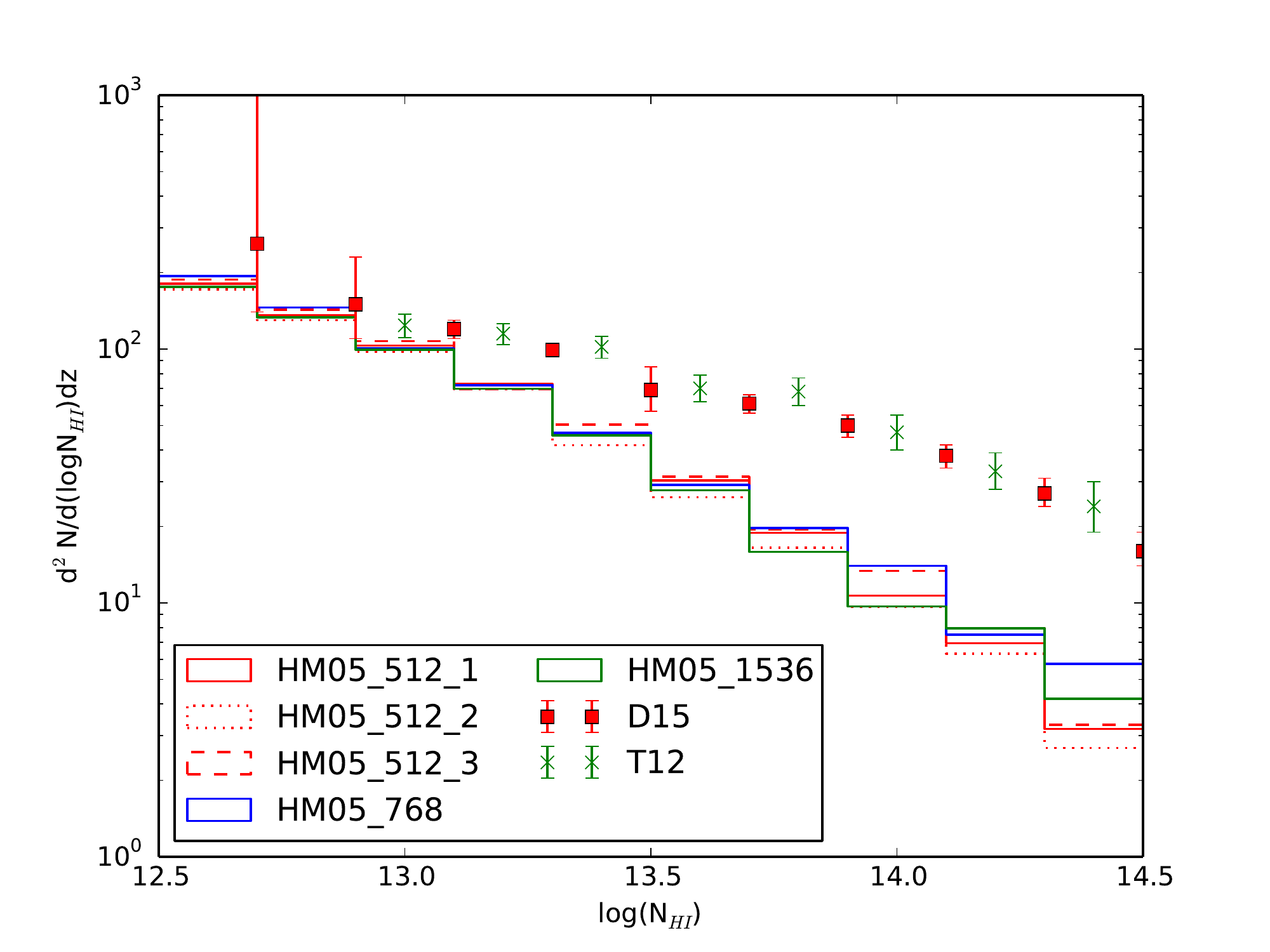}
\caption{ \small{Same format as Figure 2.  To test convergence, we compared three ($50 h^{-1}$~Mpc, $512^3$)
simulations, labeled as HM05$\_i$ (subscripts $i =$ 1, 2, 3) with two larger simulations: a new $768^3$ simulation 
(HM05$\_768$) and the $1536^3$ simulation (HM05$\_$1536) provided by Britton Smith and first presented in 
Shull \etal\ (2012).  All used the same (HM05) ionizing radiation field and are compared to \HST\ spectroscopic 
survey data from STIS (T12) and from COS (D15).} \\
} 

\end{figure}




\begin{figure}
 \epsscale{1.2}
 \plotone{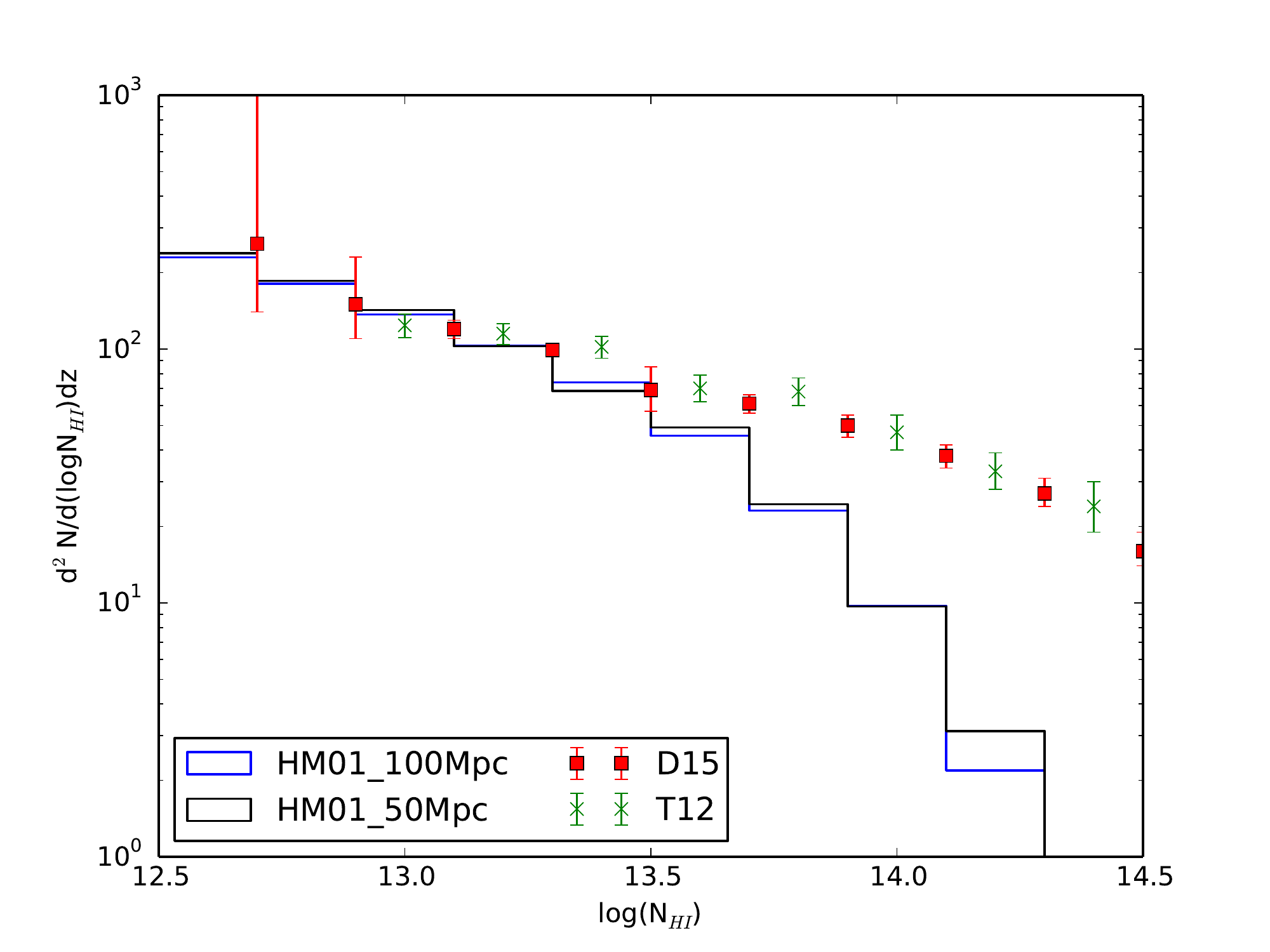}
\caption{ {\small Same format as Figure 2.  To test effects of box size, we show two simulations, 
one at ($100 h^{-1}$~Mpc, $512^3$) and another at ($50 h^{-1}$~Mpc, $256^3$), both using the
HM01 ionizing background.  Results are compared to our \HST\ survey data (D15 and T12).  
The grid cell sizes are identical ($195 h^{-1}$~kpc), and the results are essentially the same for 
$\log N_{\rm HI} < 14$.   The roll-off in simulated absorbers at $\log N_{\rm HI} > 13.7$ is an artifact 
of the poor resolution in these runs. \\  }
} 

\end{figure}




\begin{figure}
 \epsscale{1.2}
 \plotone{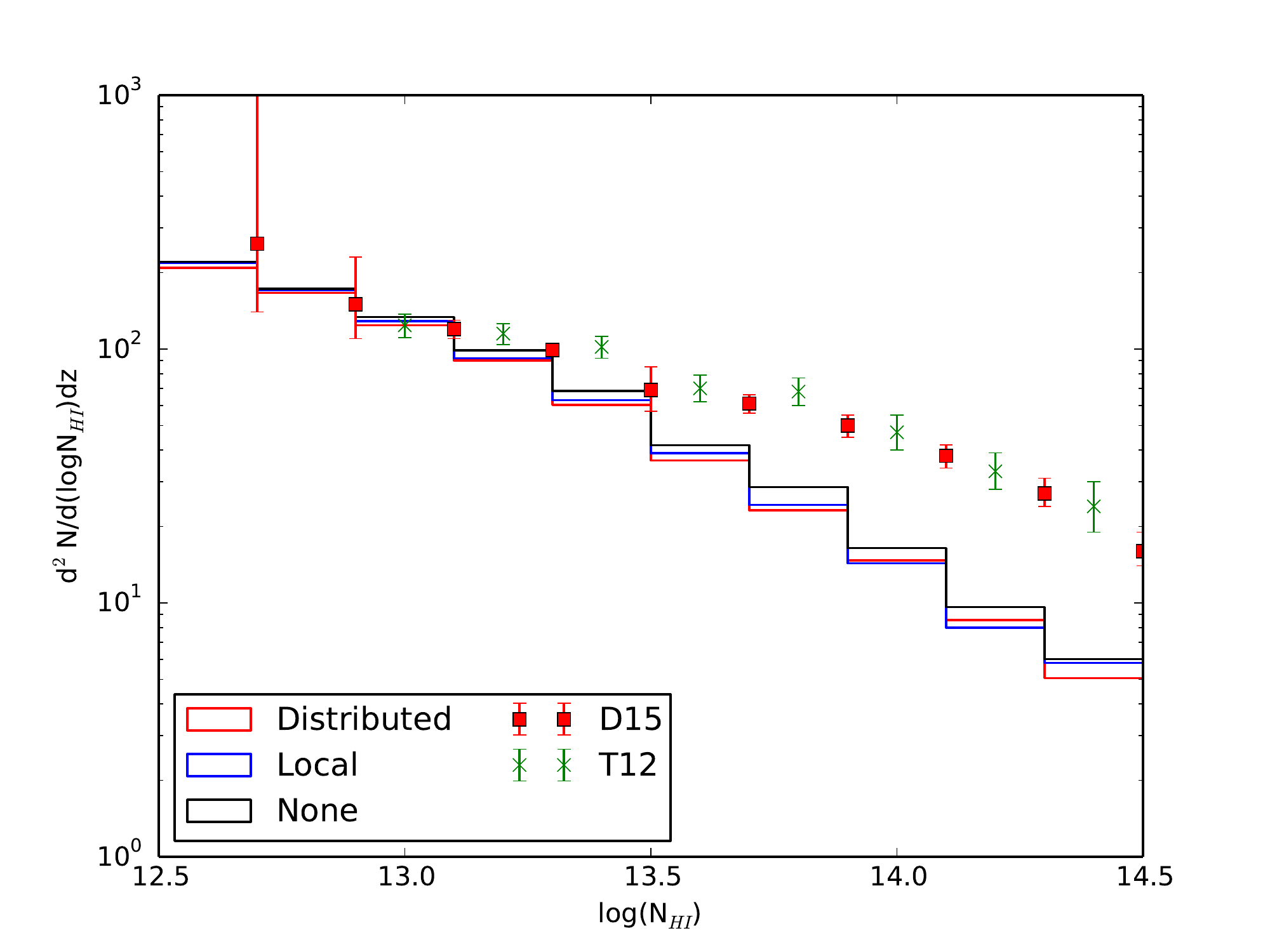}
\caption{ {\small Same format as Figure 2.  To test effects of feedback injection methods, 
we show three ($50 h^{-1}$~Mpc, $512^3$) simulations following the feedback prescriptions of 
Smith \etal\ (2011).  We explored distributed feedback (into the adjoining 27 cells), local feedback 
(into a single cell), and no feedback from star formation.    All simulations are run with the same 
HM01 radiation field and compared to \HST\ survey data (D15 and T12). \\ }
} 

\end{figure}




\begin{figure}
 \epsscale{1.2}
 \plotone{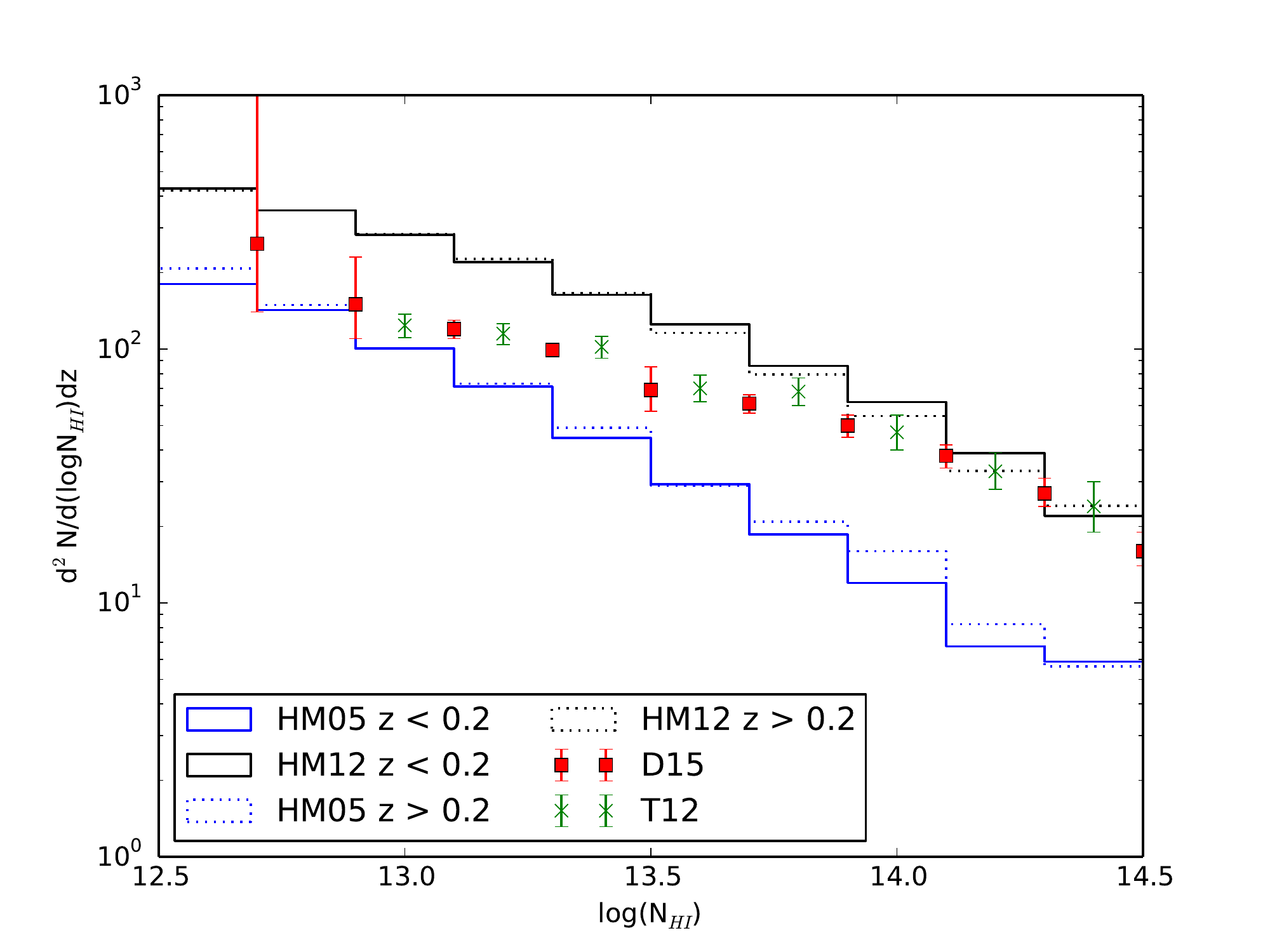}
\caption{ {\small  Same format as Figure 2.  To test for redshift evolution 
of the low-redshift \Lya\ forest, we split the results into two redshift intervals ($0 < z < 0.2$ 
and $0.2< z < 0.4$) for $768^3$ simulations, run with both HM05 and HM12 radiation fields.  
Evolution in the distribution of weak \Lya\ absorbers in the \HST\ survey (D15 and T12) are 
small, differing by less than 5\% between $0 < z < 0.4$. \\ } 
} 
\end{figure}


Figure 6 explores the potential influence of our feedback methods.  In our $512^3$ simulations, feedback has little 
effect on the distribution of \HI\ column densities. We find little difference if we inject mass, metals, and energy within 
a single cell, in adjacent cells, or eliminating feedback entirely.  Clumpiness of the IGM on small scales can be affected 
by the feedback method, even though the large-scale structural parameters, ($T, \rho_b$) and ($\Delta_b, N_{\rm HI}$),
remain nearly the same.  
Figure 7 investigates the possible effects of redshift evolution of the \Lya\ forest.  We split the simulated results into 
two redshift intervals ($0 < z < 0.2$ and $0.2< z < 0.4$) in our $768^3$ simulations, all run with the HM01 radiation
field.  Differences between the simulated distributions are quite small and consistent with COS observations
(Danforth \etal\ 2015), in which the weak \Lya\ absorbers ($\log N_{\rm HI} < 14$) exhibit little redshift evolution 
between $z = 0$ and $z = 0.4$.  

\subsection{Comparisons to Previous Simulations} 

To compare our grid-code simulations with previous SPH models, we examined two structural measures of the IGM.  The first 
measure is the thermal-phase diagram of ($T, \rho_b$), the temperature vs.\ baryon density in the low-redshift \Lya\ forest.
For our grid models, these diagrams were shown as Figure 19 in Smith \etal\ (2011) and as Figure 1 in Shull \etal\ (2012b).
For the SPH models, we examined Figure 8 of Dav\'e \etal\ (2010) and Figure 7 of Tepper-Garcia \etal\ (2012).  All of these
diagrams are consistent with a relation $T = (5000~{\rm K}) \Delta_b^{0.6}$ where $\Delta_b$ is the baryon overdensity.  
A second measure is the correlation of baryon overdensity and column density $(\Delta_b, N_{\rm HI}$), typically expressed 
as a power-law, $\Delta_b = \Delta_0 N_{14}^{\alpha}$, where $\Delta_0$ is the normalization at a fiducial \HI\ column density 
$N_{\rm HI} = (10^{14}\, {\rm cm}^{-2}) N_{14}$.   These correlations vary with redshift, but typically are evaluated at $z = 0.25$.  
Dav\'e  \etal\ (2010) used $(384^3, 48h^{-1}~{\rm Mpc})$ simulations with the HM01 background to derive a fit for absorbers 
with $T < 10^{4.5}$~K,
\begin{equation}
   \Delta_b = (35.5 \pm 0.03) N_{14}^{0.741 \pm 0.003} \, f_{\tau}^{-0.741}  \, 10^{-0.365z}  \;  .
\end{equation} 
Here, $f_{\tau} \approx 2/3$ is a renormalization factor of optical depths introduced to multiply the simulated optical depths to 
match the mean flux decrement $D_A$.  Thus, at $z = 0.25$ and with $f_{\tau} \approx 2/3$, they found
$\Delta_b = (38.9) N_{14}^{0.741}$.  From $(512^3, 100h^{-1}~{\rm Mpc})$ simulations with the HM01 background, 
Tepper-Garcia \etal\ (2012) found $\Delta_b = (48.3) N_{14}^{0.786\pm0.010}$, with no renormalization factor $f_{\tau}$.  
They noted a theoretical expectation that $\Delta_b \propto N_{\rm HI}^{0.738}$ for absorbing gas in hydrostatic and 
photoionization equilibrium (Schaye 2001).  In our $(768^3, 50 h^{-1}$~Mpc) simulations, also using the HM01 radiation field, 
we find $\Delta_b = (36.9) N_{14}^{0.65}$ over the range $0.2 < z < 0.3$ for column densities $12.5 < \log N_{\rm HI} < 14.5$.
The overdensity is calculated as a weighted average over the line elements that make up the absorber, using the column 
density as the weight field.  We find similar normalizations, $\Delta_0 = (36.9, 38.9, 48.3)$ for the three simulations 
(Shull \etal\ 2015, Dav\'e \etal\ 2010, Tepper-Garcia \etal\ 2012) evaluated at $z = 0.25$ with the HM01 background.  

Evidently, the structural measures of the low-$z$ \Lya\ forest give similar results.  The thermal-phase diagrams are 
essentially identical, while the differences in the $(\Delta_b, N_{\rm HI}$) correlation likely arise from different methods of 
identfying and characterizing H I absorbers in column density and over-density.  These comparisons are complicated by the 
fact that Dav\'e \etal\ (2010) corrected their simulated optical depths by a factor ($f_{\tau}$), and Tepper-Garcia \etal\ (2012) 
corrected for the temperature-density correlation of gas in hydrodynamic and photoionization equilibrium.  The simulations 
also used different cosmological parameters.  Tepper-Garcia \etal\ (2012) adopted density fractions 
($\Omega_m = 0.238$ and $\Omega_b = 0.0418$) from WMAP-3, which are lower by 15.6\% and 9.1\% than our WMAP-9 values 
($\Omega_m = 0.282$ and $\Omega_b = 0.046$).  Dave \etal\ (2010) adopted WMAP-7 parameters ($\Omega_m = 0.28$ and 
$\Omega_b =  0.046$) similar to our values.  Kollmeier \etal\ (2014) used $\Omega_m = 0.25$  (10\% lower than our value) and 
$\Omega_b = 0.044$ (4.3\% lower). The simulations also have different resolutions:  our standard simulations use ($768^3$, 
$50h^{-1}$ Mpc) while Tepper-Garcia use ($512^3$, $100h^{-1}$ Mpc).  Given the scatter in these correlations, their redshift 
dependence, and the correction factors, the modest differences do not warrant concern.  

 \section{SUMMARY:   A HIGHER UV BACKGROUND?}  
 
Using our recent \HST\ spectroscopic surveys of intergalactic \Lya\ absorbers, we have characterized their
distribution in \HI\ column density and redshift.   By comparing the observed distribution to new \texttt{Enzo} 
simulations of the low-redshift IGM, we find an ionizing background {\it intermediate} between the HM01 and HM12 
values.   As shown in Appendix A, the inferred baryon density and absorption-line frequency depend inversely on 
the UVB and temperature, scaling as $\Gamma_{\rm H}^{-1/2} T^{-0.363}$.  To fit the the \HST\ data, we require 
approximately a factor of 2--3 increase in the photoionization rate above HM12, somewhat less than the factor of 
five suggested by K14.  However, no single UVB agrees with the full distribution.   Figure~2 suggests that a higher 
UVB is needed to explain the line frequency of weak absorbers ($12.7 < \log N < 13.9$), while the lower HM12 
background is consistent with the distribution of stronger absorbers ($\log N_{\rm HI} > 14$).  Given the uncertainties 
in source emissivities, cosmological radiative transfer, and galaxy LyC escape fractions, our \HI\ results do not 
constitute a crisis in our understanding of the sources of the UVB.  

We have explored a number of potential explanations for the differences between the \HI\ distributions produced by
various simulations and their sub-grid feedback schemes, gaseous sub-structure, and injection of mass and metals.   
In photoionized \Lya\ absorbers, the neutral hydrogen density depends on $n_H^2$.  Consequently, the UVB and 
photoionization rate needed to explain the amplitude of $f(N_{\rm HI},z)$ may be sensitive to the clumping factor, 
$C_H \equiv \langle n_H^2 \rangle / \langle n_H \rangle ^2$.  However, a detailed code comparison is beyond the 
scope of what we can do at this time.   In the comparisons discussed earlier, we demonstrated that the integrated 
column densities, $N_{\rm HI}$, are less sensitive to the procedures for identifying absorbers or the feedback schemes.   
After investigating convergence, cosmic variance, and feedback, we conclude that the differences must arise elsewhere.  \\

The primary influence on the distribution of \HI\ absorbers in the \Lya\ forest is the ionizing radiation field.  A larger 
UVB was also found in calculations that included contributions from both quasars and galaxies (Shull \etal\ 1999; 
Faucher-Gigu\`ere \etal\ 2009).  The primary uncertainties in the UVB are the contribution from massive stars in 
galaxies (Topping \& Shull 2015) and the LyC escape fraction (HM12; Benson \etal\ 2013), a highly directional quantity 
that is difficult to constrain statistically from direct observations.  The parameterization adopted by HM12, 
$f_{\rm esc} = (1.8\times10^{-4})(1+z)^{3.4}$, was tuned to match observations at $z > 2.5$, but it is likely much too 
low at $z < 2$ where observational constraints are rare.  Reliable values require direct measurements of LyC fluxes 
from a statistically significant sample ($N_{\rm gal} \gg 20$) of starburst galaxies to constrain the low escape fractions 
expected from a highly directional LyC escape geometry.   For example, if theoretical models predict 
$\langle f_{\rm esc} \rangle \approx 0.05$, with 5\% of the LyC escaping from each side of a gaseous disk through 
perpendicular conical chimneys (Dove \& Shull 1994), direct detections would be possible,
on average, in only 5\% of the observations.   Constraints on $f_{\rm esc}$ will require large samples to deal 
with inclination bias.  Such a situation is seen in direct-detection observations at $z \approx 3$ 
(Shapley \etal\ 2006; Nestor \etal\ 2013; Mostardi \etal\ 2013) and  at $z \approx 0.2$ (Heckman \etal\ 2011) which 
find either large (10-40\%) fractions of transmitted LyC flux or none at all.   The difficulties with low-$z$ constraints 
on $f_{\rm esc}$ were discussed by Shull \etal\ (2014) in their low-$z$ census of IGM metal abundances and ionization 
ratios (C$^{+3}$/C$^{+2}$ and Si$^{+3}$/Si$^{+2}$).  \\

We therefore suggest that starburst galaxies are important contributors to the ionizing UVB at $z < 2$. Their 
contribution to the ionizing background would explain the observed distribution of \Lya\ absorbers and could resolve 
the discrepancy in IGM photoheating inferred from the opacity of the \Lya\ forest at $z < 2$ (Puchwein \etal\ 2015).  
It has also been suggested that TeV emission from blazars could add significant heat to the IGM through 
pair-production of high-energy electrons and positrons (Puchwein \etal\ 2012).  We have not investigated their 
effects on our data, either on $D_A(z)$ or the distribution in \HI\ column density.  We summarize the main results 
of our study as follows: 
\begin{enumerate}     

\item  Compared to \texttt{Enzo} N-body hydrodynamical simulations of the low-$z$ IGM with different 
    values of the ionizing UV background, the observed distribution of \Lya\ forest absorbers ($z < 0.5$)
    lies intermediate between the HM01 and HM12 background calculations.  A fit to the observations 
    requires a factor-of-two increase in the UVB above HM12 with a recommended hydrogen ionization 
    rate $\Gamma_{\rm H}(z) = (4.6\times10^{-14}~{\rm s}^{-1})(1+z)^{4.4}$.
    
 \item  The one-sided ionizing photon flux $\Phi_0 \approx 5700$~cm$^{-2}$~s$^{-1}$ at $z = 0$ agrees 
   with the observed IGM metal ionization ratios, \CIII/\CIV\ and \SiIII/\SiIV\  (Shull \etal\ 2014), and 
   suggests a 25--30\% contribution of \Lya\ absorbers to the cosmic baryon inventory. 
  
\item The increased ionizing background probably requires an increase in the escape fraction 
    of ionizing (LyC) radiation from starburst galaxies above the low values ($f_{\rm esc} < 10^{-3}$) 
    adopted by HM12.  Ionizing photons from galaxies with $\langle f_{\rm esc} \rangle \approx 0.05$ 
    gives results consistent with previous UVB modeling (Shull \etal\ 1999) that found similar contributions 
    from AGN and starburst galaxies, with specific intensities (\uvunits) at 13.6 eV of 
    $I_{\rm AGN} = 1.3^{+0.8}_{-0.5} \times 10^{-23}$ and $I_{\rm Gal} = 1.1^{+1.5}_{-0.7} \times 10^{-23}$.
   
\item Because LyC propagation through the ISM is expected to be highly directional, detections of escaping 
   photons depend on galaxy orientation and will require large surveys to obtain valid statistical inferences.  

\end{enumerate}

Future observational and theoretical work could significantly improve our characterization of the ionizing UVB 
at $z < 1$.  We need better accuracy of the column-density distribution of \Lya\ absorbers at
$\log N_{\rm HI} \leq 13.0$ and $\log N_{\rm HI} \geq 14.5$.  Larger surveys of IGM absorbers will allow us to 
measure the redshift evolution of the distribution, $d^2 {\cal N} / d(\log N_{\rm HI}) dz$, to test the expected 
increase in photoionization rate, $\Gamma_{\rm H}(z) \propto (1+z)^{4.4}$.  The most critical future experiment 
will be direct measurements of LyC escape fractions from a large sample of starburst galaxies at $z < 0.4$. 

\acknowledgments

This work was supported by NASA grant NNX08AC14G for COS data analysis and STScI archival grant 
AR-11773.01-A.   We appreciate the efforts of Britton Smith and Devin Silvia in providing the light rays 
from unpublished \texttt{Enzo} $1536^3$ models computed on the XSEDE supercomputer and used to 
construct column density distributions.  
We thank Ben Oppenheimer, Ewald Puchwein, Martin Haehnelt, Mark Giroux, and John Stocke for helpful 
discussions.   This work utilized the {\it Janus} supercomputer, operated by the University of Colorado and 
supported by the National Science Foundation (award number CNS-0821794), the University of Colorado 
Boulder, the University of Colorado Denver, and the National Center for Atmospheric Research.  Computations 
described in this work were performed using the publicly-available \texttt{Enzo} code (http://enzo-project.org),
which is the product of a collaborative effort of many independent scientists from numerous institutions around 
the world.


\clearpage


\section{Appendix~A:  Dependence of \Lya\ Absorbers on Ionization Rate}

The intergalactic  \Lya\ absorbers are thought to arise as fluctuations in dark-matter confined clumps or filaments.  
Modeling their distribution and frequency throughout intergalactic space depends on the cosmological evolution of 
this gas, coupled to calculations of the ``neutral fraction", $f_{\rm HI} \equiv n_{\rm HI}/n_H$, produced by the
balance of photoionization and radiative recombination.  These physical processes can be modeled by N-body
hydrodynamic simulations, but one can also derive their general behavior through simple analytic arguments 
(Shull \etal\ 2012b).  Here, we describe the key assumptions and parameters to illustrate how the baryon 
density and hydrogen ionization rate $\Gamma_{\rm H}$ can be inferred from observations of the \HI\ 
column density distribution and the frequency, $d{\cal N}/dz$, of absorbers per unit redshift.  

We assume a distribution of IGM absorbers with constant co-moving space density, $\phi(z) = \phi_0 (1+z)^3$,
and constant cross section $\pi p^2$.  We adopt a cosmological relation between proper distance and redshift, 
$d \ell / dz = c (dt/dz) = [c / (1+z) H(z)]$, where the Hubble parameter at redshift $z$ in a flat $\Lambda$CDM 
cosmology is $H(z) = H_0 [\Omega_m (1+z)^3 + \Omega_{\Lambda}]^{1/2}$.  In this model, the frequency of 
absorbers is $d {\cal N} / dz = \left[ c  / (1+z) H(z) \right]  \pi  p^2 \, \phi (z)$.  The internal density distributions of 
the absorbers are approximated as singular isothermal spheres, and the IGM baryon density and \HI\ fractions 
follow from photoionization models that translate the observed \HI\ to total hydrogen.  
To compare with observations, we integrate through the cloud along a chord at impact parameter 
$p = (100~{\rm kpc}) p_{100}$ of the AGN sight line through the absorber.  We adopt a 100-kpc 
characteristic scale length of \Lya\ absorbers, normalized at fiducial \HI\ column density 
$N_{\rm HI} = (10^{14}\;{\rm cm}^{-2}) N_{14}$.  Because most \Lya-forest absorbers are low density and 
optically thin in their Lyman lines, we use a case-A hydrogen recombination rate coefficient, 
$\alpha_{\rm H}^{(A)} = (4.09 \times 10^{-13}~{\rm cm}^3~{\rm s}^{-1}) T_4^{-0.726}$, scaled to temperature 
$T = (10^4~{\rm K})T_4$.  For nearly fully ionized H and He, with helium abundance 
$y = n_{\rm He} / n_{\rm H} \approx 1/12$ by number, we adopt electron density $n_e = (1+2y) n_{\rm H}$ and 
mean baryon mass per hydrogen $\mu_b = (1 + 4y) m_H$.

The co-moving baryon mass density of \Lya\ absorbers of column density $N_{\rm HI}$, probed at impact 
parameter $p$, is the product of absorber space density, 
\begin{equation}
   \phi(z) =  \frac { (d {\cal N} / dz) H(z) (1+z) } { (\pi  p^2) \, c} \; \; , 
\end{equation}  
and absorber mass 
\begin{equation}
   M_b(p) = 4 \pi \mu_b \,  p^{5/2} \left[ \frac {2 \,  \Gamma_{\rm H}(z) \, N_{\rm HI}(p) } 
      { \pi (1 + 2y ) \alpha_H^{(A)} } \right]^{1/2} \; .  
\end{equation}
The co-moving mass density, $\rho_b = \phi_0 M_b \propto [p\, N_{\rm HI} \, \Gamma_{\rm HI}] ^{1/2} (d{\cal N}/dz)$, 
is then integrated over the distribution in \HI\ column density.   We normalize $\rho_b$ to the 
closure parameter, $\Omega_b^{(\rm HI)} = \rho_{b}  / \rho_{\rm cr}$ at redshift $z = 0$, relative to critical
density $\rho_{\rm cr} = (3 H_0^2/8 \pi G)$.  Because our \HST\ surveys extend to $z \approx 0.5$, we must
include the cosmological evolution in $H(z)$, $\phi(z)$, and $\Gamma_{\rm H} (z)$.   Before introducing these 
corrections, we outline the method at low redshift, for which the Euclidean calculation gives
\begin{equation}
   \Omega_b^{(\rm HI)} = \left[ \frac {32 (2 \pi)^{1/2}}{3} \right]   \left( \frac {d{\cal N}}{dz} \right) 
       \left( \frac {G \mu_b} {c H_0} \right) \left[ \frac { p \, N_{\rm HI} \, \Gamma_{\rm H}}
              { \alpha_{\rm H}^{(A)} (1+2y) } \right] ^{1/2}     \nonumber \\
        = (7.1 \times 10^{-3}) \left[ \frac {d {\cal N} / dz} {100} \right]  \left[ p_{100} \; N_{14} 
            \left( \frac {\Gamma_{\rm H} } {\Gamma_{\rm HM12}} \right) \right]^{1/2} h_{70}^{-1}\;  T_{4}^{0.363} \; . 
\end{equation}
Here, we normalize the \Lya\ frequency to a characteristic value $d{\cal N}/dz = 100$ and scale 
$\Gamma_{\rm H}$ to the HM12 ionization rate,
$\Gamma_{\rm HM12} = (2.28 \times 10^{-14}~{\rm s}^{-1})(1+z)^{4.4}$, which we fitted over the range $0 < z < 0.7$.   
The weak temperature dependence comes from the recombination rate coefficient.  
The coefficient $(7.1\times10^{-3})$ corresponds to $\sim15$\% of the cosmological baryon fraction,
$\Omega_b = 0.0463 \pm 0.00093$ (Hinshaw \etal\ 2013; Ade \etal\ 2014).  For fixed baryon density
$\Omega_b^{\rm (HI)}$ in the \Lya\ forest, the {\it observed} frequency of \Lya\ absorbers,
$d{\cal N}/dz \propto \Gamma_{\rm H}^{-1/2}$, scaling as the inverse square root of the ionizing background.  

Our \HST\ surveys of \Lya\ absorbers provide accurate values of $\Omega_b^{\rm (HI)}$, based on an
integration over the column-density distribution. After inserting the redshift corrections to $\phi(z)$, $H(z)$, and 
$\Gamma_{\rm H}(z)$, we derive a general expression for  $\Omega_b^{(\rm HI)}$, 
\begin{equation}
 \Omega_b^{(\rm HI)} = (7.1 \times 10^{-5}) \left[ \frac {\Gamma_{\rm H}} {\Gamma_{\rm HM12} } \right]^{1/2}
      \left[ \frac {H(z)} {H_0} \right]^{1/2}  h_{70}^{-1} p_{100}^{1/2} \; T_{4}^{0.363} (1+z)^{0.2}\nonumber \\
    \times  \int_{N_{\rm min}}^{N_{\rm max}} \frac {d^2{\cal N} } {d \log (N_{\rm HI}) dz} \;  
           N_{14}^{1/2} \; d(\log \, N_{\rm HI})    \; . 
\end{equation}        
Owing to the $n_H^2$ dependence of neutral fraction in photoionization equilibrium, the factor $\Gamma_{\rm H}(z)$ 
enters Equations (8) and (10) as the square root, with a $(1+z)^{2.2}$ dependence that nearly balances the $(1+z)^2$ from the 
cosmological ratio $[\phi(z)/(1+z)$] in Equation (7).  Two recent \HST\ surveys (Tilton \etal\ 2012; Danforth \etal\ 2015) 
probed \Lya\ absorbers out to $z \approx 0.47$ and found that between 24\% and 30\% of the baryons reside in the 
low-$z$ \Lya\ forest.  Equation (10) is consistent with those estimates, particularly with our recommended factor-of-two 
increase in ionizing background, $\Gamma_{\rm H}(z) = (4.6\times10^{-14}~{\rm s}^{-1})(1+z)^{4.4}$ over the range 
$0 < z < 0.7$.

\end{document}